\documentclass[preprint2]{aastex}

\shorttitle{Evolution of Sigma Dissipationless Mergers}
\shortauthors{Stickley \& Canalizo}

\usepackage{natbib}
\usepackage{graphics}

\newcommand{\sig}{$\sigma_{\!*}$}
\newcommand{\sigd}{$\sigma_{\rm d}$}
\newcommand{\eps}{$\epsilon$}
\newcommand{\tdyn}{$t_{\rm dyn}$}
\newcommand{\msig}{$M_{\rm BH}$--$\sigma_{\!*}$}
\newcommand{\mbh}{$M_{\rm BH}$}
\newcommand{\rhalf}{$r_{\rm h}$}

\begin{document}

\title{THE EVOLOUTION OF STELLAR VELOCITY DISPERSION DURING DISSIPATIONLESS GALAXY MERGERS}

\author{\sc Nathaniel R. Stickley and Gabriela Canalizo}
\affil{Department of Physics and Astronomy, University of California,
    Riverside, CA 92521, USA}
\received{2011 September 27}
\accepted{2011 December 16}

\begin{abstract}
Using $N$-body simulations, we studied the detailed evolution of central stellar velocity dispersion, \sig, during dissipationless binary mergers of galaxies. Stellar velocity dispersion was measured using the common mass-weighting method as well as a flux-weighting method designed to simulate the technique used by observers. A toy model for dust attenuation was introduced in order to study the effect of dust attenuation on measurements of \sig.  We found that there are three principal stages in the evolution of \sig\ in such mergers: oscillation, phase mixing, and dynamical equilibrium. During the oscillation stage, \sig\ undergoes damped oscillations of increasing frequency. The oscillation stage is followed by a phase mixing stage during which the amplitude of the variations in \sig\ is smaller and more chaotic than in the oscillation stage. Upon reaching dynamical equilibrium, \sig\ assumes a stable value. We used our data regarding the evolution of \sig\ during mergers to characterize the scatter inherent in making measurements of \sig\ in non-quiescent systems. In particular, we found that \sig\ does not fall below 70\% nor exceed 200\% of its final, quiescent value during a merger and that a random measurement of \sig\ in such a system is much more likely to fall near the equilibrium value than near an extremum. Our toy model of dust attenuation suggested that dust can systematically reduce observational measurements of \sig\ and increase the scatter in \sig\ measurements.

\end{abstract}


\section{INTRODUCTION}

The central stellar velocity dispersion, \sig, of a galaxy is a key observable quantity in galactic astronomy.  Its importance is primarily due to the fact that \sig\ can be used as a proxy for the gravitational potential when interpreted using the Jeans equations \citep{Jeans1915} or the Virial theorem. In particular, there are two important galaxy scaling relations involving \sig\ which are of great importance to the study of galaxy formation and evolution: the fundamental plane (FP) of elliptical galaxies \citep{dressler1987,davies1987,bender1992} and the \msig\ relation \citep{ferrarese2000,gebhardt2000, tremaine2002}.

The FP can be expressed in various forms, most commonly $L\propto\sigma_*^{8/3} \Sigma_{\rm e}^{-3/5}$ where $L$ is the luminosity of the galaxy and $\Sigma_{\rm e}$ is the average surface brightness within the half-light radius ($R_{\rm e}$), or alternatively $R_{\rm e}\propto\sigma_*^{1.4}I_{\rm e}^{-0.9}$  where the intensity, $I_{\rm e}=I(R_{\rm e})$. The relation is significant because it can be used to estimate the distance to an elliptical galaxy, but perhaps more importantly, because its existence yields clues to how elliptical galaxies are formed. Some studies have suggested that ongoing mergers and recent merger remnants do not fit well onto the FP relation, while other studies show that late-stage mergers (i.e., mergers containing a single nucleus) can fit near or directly on the FP relation if several complicating factors are properly taken into account \citep[for detailed discussions, see][and references therein]{shier1998, bender1992, dasyra2006, rothberg2006, rothberg2011}. The two primary complicating factors that arise when studying the FP relation in ongoing and recent mergers are (1) the presence of dust obscuration and (2) enhanced star formation, which temporarily increases the relative abundance of O and B stars. Therefore, to gain a better understanding of the FP relation---especially in merger remnants---stellar dynamics must be disentangled from these complicating factors.

The \msig\ relation is a tight relation between the mass of the central black hole (\mbh) and \sig, of the form $M_{\rm BH}\propto\sigma_*^\alpha$. There is evidence that black hole (BH) activity may play a role in galaxy formation and evolution, e.g., by regulating star formation through winds and outflows. To understand galaxy formation, we therefore need to understand how super-massive BHs and galaxies co-evolve. The \msig\ relation is one of the most important tools available in this endeavor. In order to fully understand the origin and implications of the \msig\ relation, we must determine how the relation evolves with cosmological time (i.e., redshift). Active galactic nuclei (AGN) have become instrumental in studying the evolution of the \msig\ relation with redshift because AGN hosts are the only galaxies for which we can measure \mbh\ in the non-local universe. Unfortunately, limiting the study of \msig\ to AGN hosts may introduce a bias or significantly increase the observed scatter.  This is because AGN activity has often been linked with galaxy merger activity \citep{canalizo2001}. Many modern numerical simulations that successfully reproduce present-day properties of early-type galaxies, such as the color-magnitude or the \msig\ relations, presuppose that AGN activity is triggered by mergers of gas-rich galaxies \citep{hokins2006,robertson2006a,robertson2006b}. These gas-rich mergers also trigger star formation. Star formation, in turn, adds dust to the interstellar medium; the dust attenuates star light. Therefore, in order to understand the \msig\ relation for these objects, as well as any other objects which are not dynamically relaxed, we need a better understanding of the evolution of \sig\ during the merger process as well as an understanding of how star formation and dust attenuation may influence this measurement.

In the present paper, we have taken the first steps toward a theoretical understanding of the evolution of \sig\ \emph{during} mergers of galaxies.  We have also made initial steps toward understanding how dust attenuation might influence the measurement of \sig.  Traditionally, simulation work involving \sig\ has involved making measurements of \sig\ after the merger is complete \citep[e.g.,][]{cox2006,robertson2006a, robertson2006b}. We only know of one study in which \sig\ was plotted as a function of time \textit{during} a merger \citep{johansson2009} and, in that case, the measurement interval for \sig\ was 200 Myr, which (by the Nyquist sampling theorem) implies that the simulations could only capture the presence of fluctuations in \sig\ having a period of 400 Myr or more. To put this in context, we note that the dynamical timescales of the systems in these simulations were less than 100 Myr. Furthermore, in simulations, \sig\ is often based on the velocities of the stars falling within the half-mass radius, $r_{\rm h}$, rather than on the velocities of stars appearing in a slit placed across the center of the system, as is the case in observational measurements of \sig\ (a notable exception is the work of \citet{cox2006}, which \textit{did} employ a slit). The value of \sig\ that is typically reported is the mean value measured along 100-250 random lines of sight; the standard deviation of \sig\ over the set of viewing directions is typically not reported. Finally, to our knowledge, no simulation study has attempted to measure the effect of dust attenuation on determinations of \sig. The work presented here is different in all of these regards. In our simulations, the mass-weighted and flux-weighted \sig\ were measured based upon the velocities of all stars appearing within a simulated diffraction slit centered on the nucleus of the galaxy.  The flux-weighted measurement incorporated a toy model for dust attenuation. We measured \sig\ in this way along $10^3$ lines of sight and performed a statistical analysis on the directional distribution of \sig. We were particularly interested in identifying the time-variation of \sig\ during the merger process, thus these measurements were performed at very short time intervals during the merger.  

The paper is organized as follows.  In Section 2, we describe the numerical simulations that were performed as well as the automated analysis algorithm that was used. In Section 3, we present the results of the numerical simulations and perform some additional analysis.  In Section 4, we conclude by discussing our findings, their implications, and their limitations.

\section{NUMERICAL SIMULATIONS}

In order to study the evolution of \sig\ with high time resolution during the merger process, we designed a simulation code that performs a statistical analysis of the particle velocity data at short intervals during run time. We chose to keep the simulations as simple as possible so that we could identify the purely dynamical aspects of the evolution of \sig. The simulated galaxies were composed only of gravitationally bound, collisionless star particles. There was no separate dark matter component nor a gas component. At fixed intervals, the code computed mass-weighted and flux-weighted values of \sig\ along $10^3$ random lines of sight. The measurements of \sig\ were based on the velocities of stars appearing in a rectangular ``diffraction slit'' centered on the projected center of mass of the system. This method allowed all stars along the line of sight to contribute to the measured value of \sig---just as in the case of an observational measurement of \sig. Flux-weighted values of \sig\ were determined from the intrinsic luminosities of star particles, coupled with a toy model for dust attenuation. Using the values of \sig\ measured along $10^3$ directions, the code computed the mean, minimum, maximum, standard deviation, skewness, and kurtosis of \sig\ during the merger process. 

\subsection{Code Description} 

\subsubsection*{Dynamics}

Our $N$-body simulation code utilized the direct (i.e., all pairs) method to calculate the gravitational forces on each particle. The particles were softened in order to minimize the effects of two-body relaxation \citep{white1978}. This ensured that the stellar systems were collisionless for the duration of the simulations. The acceleration of each particle, $i$, was computed using a truncated Plummer softening scheme, given by 
\begin{displaymath}
  \ddot{\mathbf{{r}}}_{i}=
     -G\sum_{j\neq i}m_{j}\mathbf{r}_{ij}\times\left\{
	\begin{array}{ll}
	    r_{ij}^{-3} & r_{ij}>\epsilon \\
	    \left(r_{ij}^2+\epsilon^2\right)^{-3/2} &  r_{ij}\leq\epsilon
	 \end{array}
      \right.
\end{displaymath}

\noindent
where $\mathbf{r}_{ij}=\mathbf{r}_{i}-\mathbf{r}_{j}$ and \eps\ is the softening length. Using this scheme, particles interact as Plummer spheres when overlapping significantly and as point particles otherwise. 
The system was integrated forward in time using the kick-drift-kick form of the leap-frog integrator \citep{quinn1997} with global, adaptive time stepping.

\subsubsection*{Galaxy Construction}

 In constructing progenitor galaxies, our code first initialized the stellar population. A list of stellar masses was made with relative abundances determined by an approximation to the Kroupa initial mass function \citep[IMF;][]{kroupa2001}. Stars with masses $M \geq 3 M_\sun$ were then removed from the list in order to simulate the effect of an aged stellar population. Each of the $N$ gravitationally softened particles in the simulation represented a population of $n$ stars, thus the mass of each particle was determined by randomly selecting $n$ entries from the stellar population list. The luminosity of each particle was determined from the sum of the luminosities of the $n$ constituent star particles, which were, in turn, computed from the mass--luminosity relation \citep{salaris}. 

Once the particles were initialized, they were distributed in space according to the Hernquist density profile \citep{hernquist1990},
\begin{equation}
 \rho(r) = \frac{M}{2\pi}\frac{D}{r(r+D)^3}
\end{equation}
\noindent
where $M$ is the total mass of the system, and the size scale, $D$, of the distribution is the ``Hernquist radius.''  The velocity distribution was chosen to reproduce the analytic result obtained from the Jeans equations, assuming an irrotational system with an isotropic dispersion tensor and Hernquist density profile (see the Appendix). The resulting system, having been constructed stochastically, exhibited a small net linear momentum, implying a small drift velocity. All particle velocities were adjusted to remove the net linear momentum. Our newly constructed system was then evolved forward in time. The density profile and central velocity dispersion of the newly constructed system varied for approximately 2.5 dynamical timescales before reaching a steady configuration. This period of adjustment was apparently necessitated by the mismatch between the smooth analytic distribution used to initialize the system and the granular approximation that was actually produced. The final density profile of the progenitor system fit the Hernquist profile perfectly---within the limits of particle noise. After the initial 2.5 dynamical timescales, no evolution was observed during the subsequent $\approx 290$ dynamical times for which the system was studied. Based on this analysis, we evolved each progenitor spheroid passively for 3.0 dynamical times before using it in a merger simulation.

\subsubsection*{Dust Attenuation}

In order to determine how measurements of \sig\ might depend on dust attenuation, a toy model for dust attenuation was included in the code. A cylindrical slab of gray attenuating material of radius, $r$, thickness, $\delta$, and extinction coefficient, $\kappa$ was placed concentric with the center of mass of the merging system. The flux of the $i$th star in the system, $f_i$, was then calculated according to
\begin{equation}
  f_i=L_ie^{-\kappa d_i},\label{atten}
\end{equation}
where $L_i$ is the luminosity of the $i$th star particle and $d_i$ is the distance that light from the $i$th star traveled through the attenuating slab on its way to the virtual observer.

\subsubsection*{Measuring \sig} 

The process used to measure \sig\ in our code resembled the common observational technique, in that a rectangular slit of width $w$ and length $\ell$ was centered on the system and the measurement of \sig\ was based on the line-of-sight components of the velocities of stars within the slit.  This method is different from the one often employed in many numerical simulations; simulators typically base \sig\ measurements on the velocities of stars within the half\-mass radius or projected half-mass radius of the system \citep{johansson2009,robertson2006a,robertson2006b}. 

The measurement of \sig\ proceeded as follows.  First, a random viewing direction ($\theta, \phi$) was chosen, then the system was rotated such that the new $+z$-axis coincided with the ($\theta, \phi$) direction. The set of particles within the observing slit was identified using the new $x$- and $y$-coordinates. The line-of-sight velocity of each particle was then given by the new $z$-component of velocity. From the line-of-sight velocity distribution within the slit, the mass-weighted and flux-weighted versions of \sig\ were computed as follows
\begin{eqnarray}
\mbox{m}\sigma_* &=& \sqrt{v_i^2m_i/M - (v_im_i/M)^2} \\
\mbox{f}\sigma_* &=& \sqrt{v_i^2f_i/F - (v_if_i/F)^2}, 
\end{eqnarray}
\noindent
with 
\begin{displaymath}
 M=\sum_i m_i \quad\quad F=\sum_i f_i,
\end{displaymath}
\noindent
where the standard summation convention has been utilized; repeated indices imply a sum over that index. We use the notation m\sig\ and f\sig\ to denote mass-weighted and flux-weighted \sig, respectively. This process was repeated for $10^3$ random directions.  

\subsubsection*{Directional Statistics}

Once m\sig\ and f\sig\ were measured for $10^3$ directions, statistical quantities were computed in order to determine the degree of anisotropy of the merger system. Specifically, the mean, minimum, maximum, standard deviation, skewness, and kurtosis of m\sig\ and f\sig\ were computed for the set of directions. Since the definitions of skewness and kurtosis vary among authors, we present the definitions that we used below
\begin{eqnarray}
\mbox{skew}(x) &=&  n^{-1}\sum_i (x_i-\mu)^3/\sigma^3 \\
\mbox{kurt}(x) &=&  n^{-1}\sum_i (x_i-\mu)^4/\sigma^4 
\end{eqnarray}\noindent
where $\mu$ and $\sigma$ are the mean and standard deviation of $x$, respectively and $n$ is the sample size. The directional statistics, combined with time-evolution data, enabled us to estimate the intrinsic scatter in observations of \sig\ for randomly oriented active mergers and merger remnants.

\subsubsection*{Collisions}

The code was specifically designed to study the simplest type of mergers: binary mergers of equal-mass systems (i.e., 1:1 mergers).  Once a model galaxy was constructed, as described above, the galaxy was replicated and placed on a collision course with its clone. The relative position and velocity vectors ($\mathbf{R}$ and $\mathbf{V}$) were specified, then the code adopted the zero-linear-momentum reference frame and set the origin of the coordinate system coincident with the center of mass. This ensured that the final merger remnant would be centered at the origin.

\subsection{Precision} \label{precision}

Particle noise was by far the main source of uncertainty in our measurement of \sig. We quantified the precision in the primary measurement of interest---the directional mean of \sig---through experimentation. First, we noted that our newly constructed galaxies were perfectly spherically symmetric and isotropic---except for the statistical noise introduced by using a finite number of particles, $N$. In the limit as $N\rightarrow\infty$, the measurement of \sig\ should be independent of the direction from which the measurement was made. This implies that the standard deviation of \sig\ over the set of all viewing directions (\sigd) should vanish in this limit. Based on this insight, we performed measurements of the standard deviation of \sig\ in newly constructed systems, measured from $10^3$ random directions, for various $N$ ranging from $10^2$ to $10^5$. We found the expected behavior: $\sigma_{\rm d}\propto N^{-1/2}$. Upon determining the constant of proportionality associated with our simulation parameters (namely, the system density and slit dimensions), we were able to quantify the degree of particle noise in our measurements of \sig.

The uncertainty introduced by the error in the numerical integration scheme was negligible in comparison with the particle noise. All non-integer quantities in the code were stored as double precision floating point numbers. We used a very stringent step-size criterion in which no particle was allowed to move more than 1.1\eps\ during a time-step. The total momentum of the system was conserved to within a small multiple of the machine precision and the total energy fluctuated by less than 0.1\%.    
  
\subsection{Simulation Units \&\ Parameters}\label{param-section}

Internally, the code used a system of units in which mass was measured in solar masses, distance was measured in parsecs, the gravitational constant $G=1$, and time was a derived unit. However, in order to make our results easier to interpret, we have re-scaled the simulation parameters. All quantities will be presented in terms of the characteristics of the progenitor galaxies. Our unit of distance is the half-mass radius, \rhalf, of the progenitor, which can be written in terms of the Hernquist radius, $D$,  as $r_{\rm h}=(1+\sqrt{2})D$. For a system with constant mass-to-light ratio, the projected half-light radius, $R_{\rm e}$, is related to \rhalf\ by $r_{\rm h} \approx 1.33 R_{\rm e}$ \citep{hernquist1990}. Our unit of time is the dynamical time-scale of the progenitor, given by
\begin{equation}
 1\, t_{\rm dyn}=\sqrt{\frac{3\pi}{16G\bar{\rho}_{\rm h}}}=\sqrt{\frac{\pi^2(1+\sqrt{2})^3D^3}{2GM}}\label{tdyn}
\end{equation}
\noindent
where $\bar{\rho}_{\rm h}$ is the mean density within \rhalf.  The unit of velocity is then given by 
\begin{equation}
 1\, \frac{r_{\rm h}}{t_{\rm dyn}}=\sqrt{\frac{2GM}{\pi^2(1+\sqrt{2})D}}\label{vunit}
\end{equation}
\noindent
where $M$ is the total mass of the system. In terms of these units, the gravitational constant is, 
\begin{equation}
 G=\frac{\pi^2}{2}\frac{r_{\rm h}^3}{M\,t_{\rm dyn}^2}\label{bigG}
\end{equation}
To gain a better understanding of this unit system, consider a spheroid of mass $1.0\times10^{10}\,\mbox{M}_\sun$ with an effective radius of $R_{\rm e} = 2.0\, \mbox{kpc}$.  For this spheroid, the time unit is $37\,\mbox{Myr}$ and the velocity unit is $70\,\mbox{km s}^{-1}$.
 
After studying a large variety of initial orbital parameters, we found that the general behavior of all of the merger simulations fell between two extreme cases: head-on collisions and orbital decay mergers.  In a head-on collision, the progenitors are initially separated by some distance, $R$, and given an initial relative speed, $V$, directly toward one another. In the orbital decay scenario, the two galaxies begin on circular or nearly circular orbits. As they interact through tidal forces and dynamical friction, angular momentum is redistributed and the galaxies gently merge. We analyzed these two extreme cases, as well as a representative intermediate case, in detail. For all simulations, the radius, thickness, and attenuation coefficient of the cylindrical attenuating slab were $r= 3.36 $ \rhalf,  $\delta = 1.24 $ \rhalf, and $\kappa= 1.8 $ $r_{\rm h}^{-1}$, respectively. The slit width and length were $w = 0.14 $ \rhalf\ and $\ell = 0.70 $ \rhalf, respectively. Table \ref{params} summarizes the parameters that were varied in our simulations.

\begin{table*}
\begin{center}
\caption{Summary of Parameters Used\label{params}}
\begin{small}
\begin{tabular}{lcccccl}
\tableline
\tableline
\vspace{-3mm}\\
Simulation Type 	& 
$N$ 			& 
\eps\ ($r_{\rm h}$)	&
$V$  ($r_{\rm h}t_{\rm dyn}^{-1}$)& 
$R$  ($r_{\rm h}$)	&
$\theta$  (deg)	&
$\tau$  ($t_{\rm dyn}$)	
\\
\tableline \vspace{-3mm}\\
Head-on, short 		& 50,000  & 0.0233 & 1.55 & 1.66 & 180	& 0.00942\\
Head-on, long 		& 35,000  & 0.0262 & 1.55 & 1.66 & 180	& 0.0942\\
Intermediate, short 	& 50,000  & 0.0233 & 1.60 & 1.69 & 153.2 & 0.00942\\ 
Intermediate, long 	& 35,000  & 0.0262 & 1.60 & 1.69 & 153.2 & 0.0942\\ 
Orbital decay, short 	& 50,000  & 0.0233 & 1.55 & 1.66 & 90 	& 0.00942\\
Orbital decay, long 	& 35,000  & 0.0262 & 1.55 & 1.66 & 90 	& 0.0942\\
\tableline 
\end{tabular}
\end{small}
\end{center}
	\tablecomments{The qualifiers ``short'' and ``long'' refer to the duration of the simulation and the interval between measurements of \sig. All ``short'' simulations were evolved forward for $\approx 8$ \tdyn\ while the ``long'' were allowed to evolve for $\approx 290$ \tdyn. The number $N$ is the number of particles in each progenitor system; the \textit{total} number of particles in the merger simulation is $2N$. The initial relative speed of the galaxies is $V=|\mathbf{V}|$, their initial separation distance is $R=|\mathbf{R}|$, and the angle between $\mathbf{R}$ and $\mathbf{V}$ is $\theta$. The interval between measurements of \sig\ is $\tau$.}
\end{table*}

\section{RESULTS}

\subsection{Measurement Techniques}

\subsubsection*{Slit \sig\ versus Half-mass \sig}

Using our slit-based method of measuring \sig, with the slit dimensions given in Section \ref{param-section}, we found that \sig\ in our progenitor systems was $\sigma_*=1.07\pm 0.01\,\,r_{\rm h}\,t_{\rm dyn}^{-1}$ while the value of \sig\ measured within the central spherical region of radius \rhalf\ was $1.0047\pm0.0026 \,\,r_{\rm h}\,t_{\rm dyn}^{-1}$. For comparison, the result obtained by analytically computing the mass-weighted mean of the line-of-sight velocity dispersion within a sphere of radius \rhalf\ for an isotropic Hernquist profile is $\approx 1.0035\,\,r_{\rm h}\,t_{\rm dyn}^{-1}$ while the same quantity computed within $r_{\rm h}/2$ is $\approx1.0693\,\,r_{\rm h}\,t_{\rm dyn}^{-1}$ (see the Appendix). Evidently, measuring \sig\ using a narrow slit placed on the center of the galaxy yielded values that were closer to the velocity dispersion within a region smaller than \rhalf. As the width of the slit was increased, the difference between the two measurement techniques diminished. In quiescent, isotropic systems, the two measurement techniques agreed within 2\% when the width of the slit $w=r_{\rm h}$ and the methods agreed to within the measurement uncertainty when $w=2r_{\rm h}$. As the width of the slit approached zero, particle noise made the measurement highly unreliable. For non-quiescent and non-isotropic systems, the two methods sometimes differed significantly---even when the slit was quite large. This disagreement was due to the motion of particles lying outside of the spherical half-mass region, which were included in the slit-based measurement, but absent from the spherical half-mass measurement. 

\subsubsection*{Flux Weighting vs. Mass Weighting}

The presence of an attenuating slab placed concentric with the galaxy had the effect of reducing the flux received from stars in the central region of the galaxy. Using the properties of the attenuating slab presented in Section \ref{param-section}, along with Equation (\ref{atten}), we see that the flux received from a star particle at the center of the galaxy was $f=0.3275L$ when viewed through the thinnest portion of the slab (face-on), where $L$ is the intrinsic luminosity of the star particle. When viewed through the thickest portion of the slab (edge-on), $f=0.0025L$. When velocity dispersions were computed using the flux-weighting technique, the velocities of stars in the central region of the galaxy were therefore weighted less heavily. Since the stellar velocity dispersion was largest near the central region of our simulated galaxies, the flux-weighted measurement, f\sig, was smaller than its mass-weighted counterpart, m\sig. Using a simulation with $10^5$ particles, we found that f$\sigma_*=(0.875\pm0.015)\mbox{m}\sigma_*$. In general, increasing the attenuation coefficient or the dimensions of the attenuating slab caused f\sig\ to decrease monotonically. 

The presence of the cylindrical attenuator also destroyed the spherical symmetry of the system; when the mass-weighted velocity distribution was isotropic, the flux-weighted velocity distribution was not. As mentioned in Section \ref{precision}, the standard deviation of \sig\ over the set of random viewing directions (\sigd) can be used as a measure of the anisotropy of the system's velocity distribution. Using the same $10^5$ particle simulation mentioned above, the standard deviation of f\sig\ over the set of $10^3$ random viewing directions was f$\sigma_{\rm d}=(0.123\pm0.017)\sigma_*$ while m$\sigma_{\rm d}=(0.006\pm0.001)\sigma_*$.  

Additionally, we found that changing the radius-to-thickness ratio of the attenuating slab ($r/\delta$) caused the anisotropy of the flux-weighted velocity distribution, f\sigd\ to change. The minimum value of f\sigd\ was found when $r/\delta\approx0.5$, of course f\sigd\ also tended toward zero if either $\delta/r_{\rm h}\rightarrow0$ or $r/r_{\rm h}\rightarrow0$, regardless of the ratio $r/\delta$. The value of $r/\delta$ leading to maximum isotropy occurred in the interval $1<r/\delta<\infty$; the exact ratio $r/\delta$ that maximized f\sigd\ depended on the relative size of the slab compared with that of the galaxy.

\subsection{Merger Evolution}

The dissipationless mergers that we studied proceeded in three primary stages: \textit{oscillation} then \textit{phase mixing}, and finally completion or \textit{dynamical equilibrium}. Figures \ref{headon-osc}--\ref{decay-osc} respectively illustrate the early stages of the ``Head-on, short'', ``intermediate, short'', and ``orbit-decay, short'' simulations from Table \ref{params}.

\subsubsection*{Oscillation}

\begin{figure*}
 \begin{center}
    \includegraphics[width=5.5in]{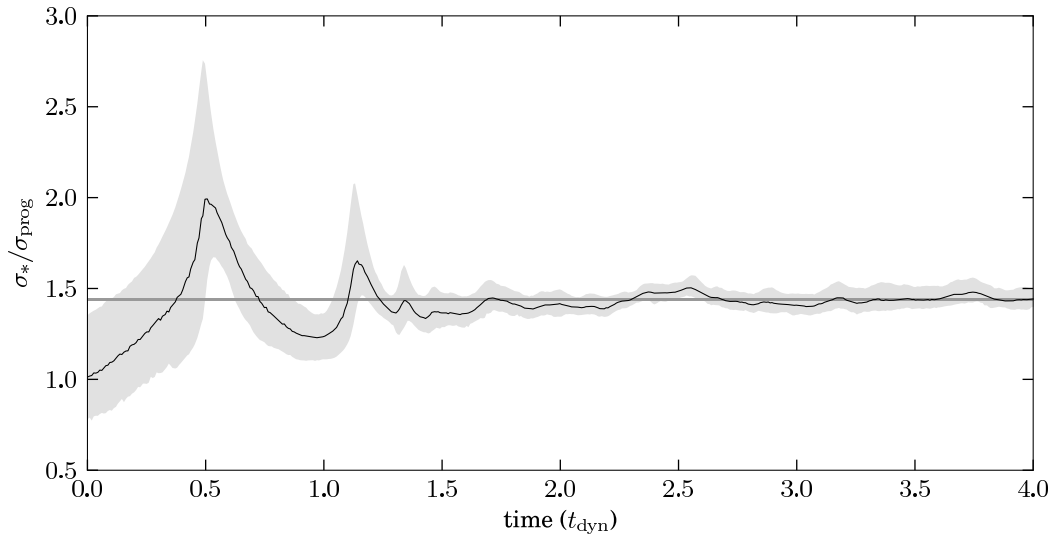}
    \caption{\label{headon-osc}\small Evolution of \sig\ with time during the ``Head-on, short'' merger of Table \ref{params}. We have scaled \sig\ by the value of velocity dispersion of the progenitor galaxies ($\sigma_{\rm prog}$). The black curve represents the directional mean of the mass-weighted line-of-sight velocity dispersion (m\sig) measured using the slit-based method. The upper and lower limits of the gray region are, respectively, the maximum and minimum values of m\sig\ for the set of $10^3$ random viewing directions. The gray horizontal stripe is centered on the final equilibrium value of m\sig\ of the merger remnant. The half-thickness of the stripe illustrates the 1$\sigma$ uncertainty due to particle noise. Thus the gray stripe is  $(\mbox{m}\sigma_{*,{\rm final}}\pm\sigma_{\rm d,noise})/\sigma_{\rm prog} = 1.440\pm0.008$. The centers of the two progenitor galaxies first coincide at $t\approx0.5$ \tdyn. The oscillation stage ends at $t\approx1.5$ \tdyn. The velocity dispersion continues to fluctuate, although less significantly, in the phase mixing stage. Note: Although this simulation was evolved forward for $\approx 8$ \tdyn, the plot only shows the first 4 \tdyn\ in order to highlight the oscillatory stage of the merger.}
 \end{center}
\end{figure*}

\begin{figure*}
     \begin{center}
     \includegraphics[scale=1]{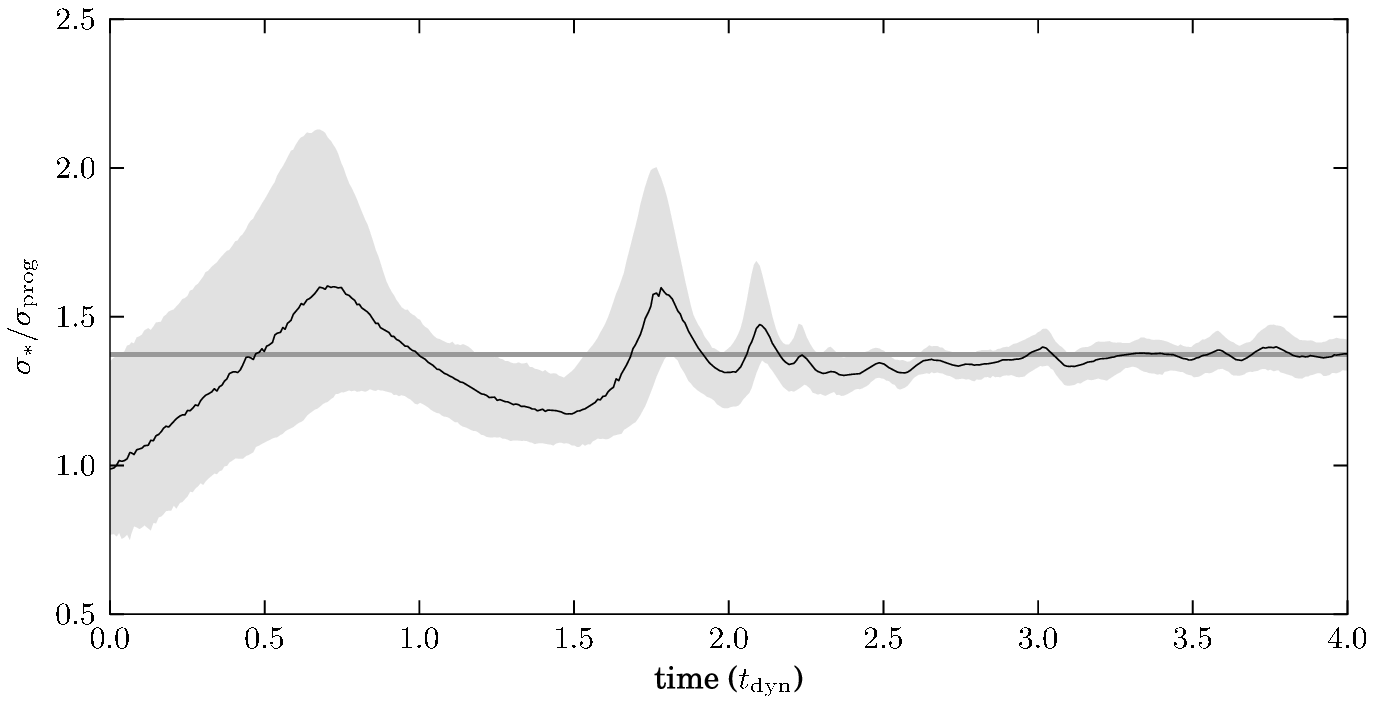}
     \caption{\label{intermediate-osc}\small Evolution of \sig\ with time during the ``Intermediate, short'' merger of Table \ref{params}. The black line, shaded region, and gray horizontal stripe are defined the same as in Figure \ref{headon-osc}. The position and thickness of the gray stripe are $(\mbox{m}\sigma_{*,{\rm final}}\pm\sigma_{\rm d,noise})/\sigma_{\rm prog} = 1.373\pm0.008$. The centers of the two progenitor galaxies first coincide at $t\approx0.7$ \tdyn. The oscillation stage ends at $t\approx2.3$ \tdyn. }
     \end{center}
\end{figure*}

\begin{figure*}
    \begin{center}
    \includegraphics[scale=1]{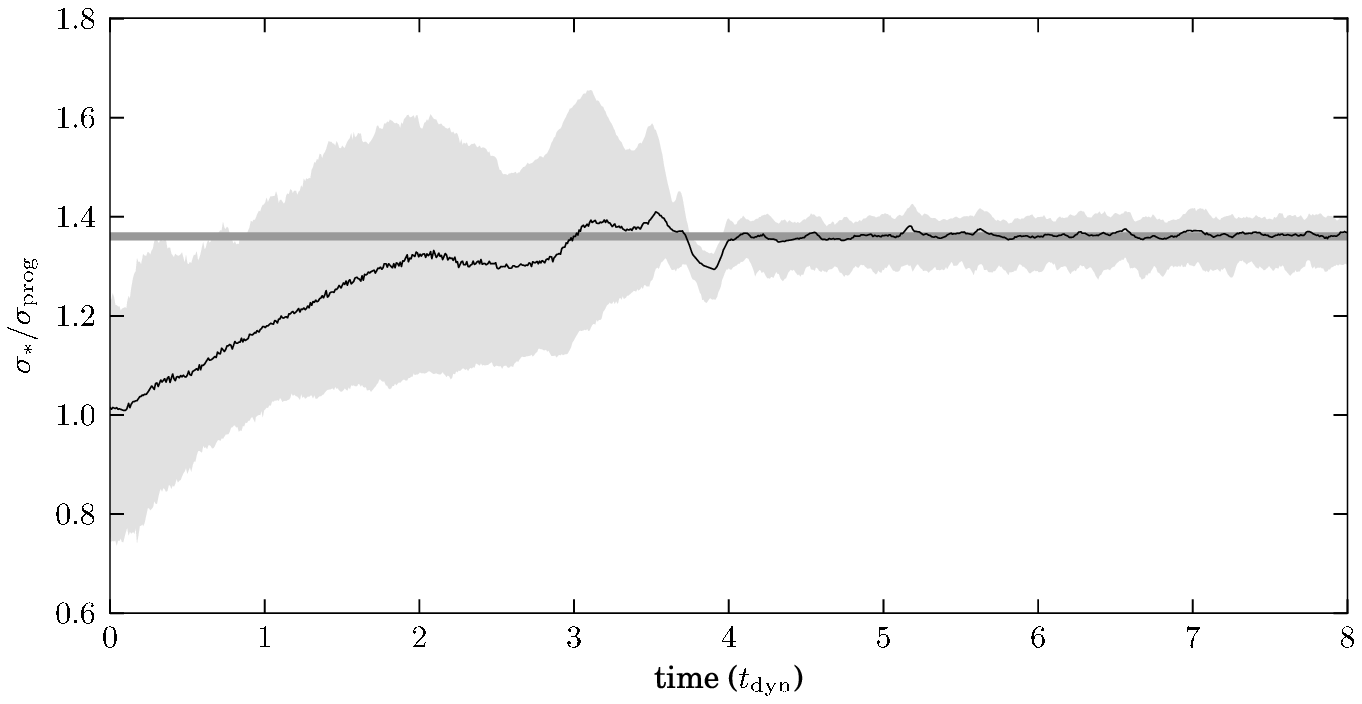}
    \caption{\label{decay-osc}\small Evolution of \sig\ with time during the ``Orbit decay, short'' merger of Table \ref{params}. The black line, shaded region, and gray horizontal stripe are defined the same as in Figure \ref{headon-osc}. The position and thickness of the gray stripe are $(\mbox{m}\sigma_{*,{\rm final}}\pm\sigma_{\rm d,noise})/\sigma_{\rm prog} = 1.361\pm0.008$. The centers of the two progenitor galaxies first coincide at $t\approx3.1$ \tdyn. The oscillation stage ends at $t\approx4.0$ \tdyn.}
    \end{center}
\end{figure*}

 The oscillatory stage is characterized by the bulk motion of the two progenitor galaxies as they coalesce. As the progenitor nuclei become superimposed significantly for the first time, the density, gravitational potential, and \sig\ increase rapidly. The progenitor nuclei then typically pass through one another. The density, gravitational potential, and \sig\ of the central region of the merging system are then temporarily reduced.  The nuclei of the systems eventually change directions and fall back onto one another, while some of the star particles that were initially less tightly bound continue on their original paths---only slightly perturbed by the motion of the nuclei. The process then repeats several times until the coherent oscillations decay away. The evolution of \sig\ then becomes dominated by phase mixing. In the Head-on merger shown in Figure \ref{headon-osc} and the Intermediate merger shown in Figure \ref{intermediate-osc}, the transition between the oscillatory and phase mixing stages occurred at approximately 1.5 \tdyn\ and 2.3 \tdyn, respectively. In the less violent Orbital decay merger shown in Figure \ref{decay-osc}, the oscillatory stage was much less pronounced because the galaxies gently spiraled into one another, but oscillations were still visible between 2.0 \tdyn\ and 4.0 \tdyn.

\subsubsection*{Phase Mixing}

The phenomenon of phase mixing arises because of the dispersion in the oscillation (or orbital) periods of the particles composing a system. Due to this dispersion, an initially coherent oscillation becomes incoherent with time and eventually vanishes. For a system consisting of a finite number of particles, this sort of phase mixing \textit{alone} cannot be fully responsible for the termination of the oscillation stage described above because the differences between the orbital periods are necessarily finite. A finite system of uncoupled, undamped, undriven oscillators will periodically return to its initial state, meaning that episodes of coherent oscillation would recur periodically.  Phase mixing plays an important role in ending the oscillation stage, but \textit{particle interactions} are essential in preventing the recurrence of coherent oscillations. The stars in a galaxy interact indirectly through a time-varying global potential. This sort of interaction, which is most obvious during the oscillatory stage of a merger, is known as ``violent relaxation'' \citep{lyndenbell1967}. Stars also interact through close encounters with other stars, however this mode of interaction is insignificant compared with violent relaxation for most stars in a typical galaxy during the timescale of a merger. These interactions cause the period of each particle to vary with time in a non-periodic way which modifies the pure phase-mixing mechanism and prevents the return of coherent oscillations.

Visually, the process of phase mixing causes the particles of the two progenitor systems in a merger to lose their identities; it gradually becomes impossible to distinguish the distribution of particles that initially belonged to progenitor A from the distribution of particles that originally belonged to progenitor B. This is illustrated particularly clearly in figure 5 of \citet{funato1992a}.

\begin{figure*}
 \begin{center}
    \includegraphics[width=5.5in]{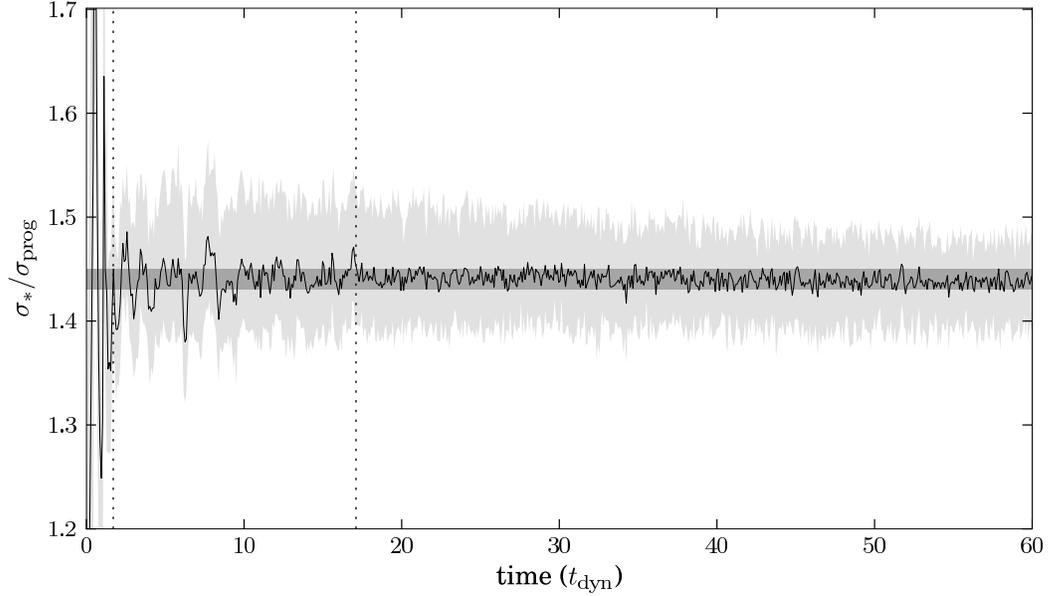}
    \caption{\label{headon-osc-long}\small The evolution of \sig\ with time during the ``Head-on, long '' merger of Table \ref{params}. The black line, shaded region, and gray horizontal stripe are defined the same as in Figure \ref{headon-osc}. The position and thickness of the gray stripe are $(\mbox{m}\sigma_{*,{\rm final}}\pm\sigma_{\rm d,noise})/\sigma_{\rm prog} = 1.4399\pm0.0096$. The phase mixing stage ends at $t\approx17$ \tdyn. The vertical dotted lines mark the approximate boundaries between the oscillation, phase mixing, and dynamic equilibrium stages. Note that the maximum value of \sig\ (the upper boundary of the shaded region) continues to evolve until $\approx 45$ \tdyn. Although this simulation was evolved forward for $\approx 290$ \tdyn, the plot only shows the first 60 \tdyn\ because there was no visible evolution beyond 60 \tdyn.}  
 \end{center}
\end{figure*}

\begin{figure*}
     \begin{center}
     \includegraphics[scale=1]{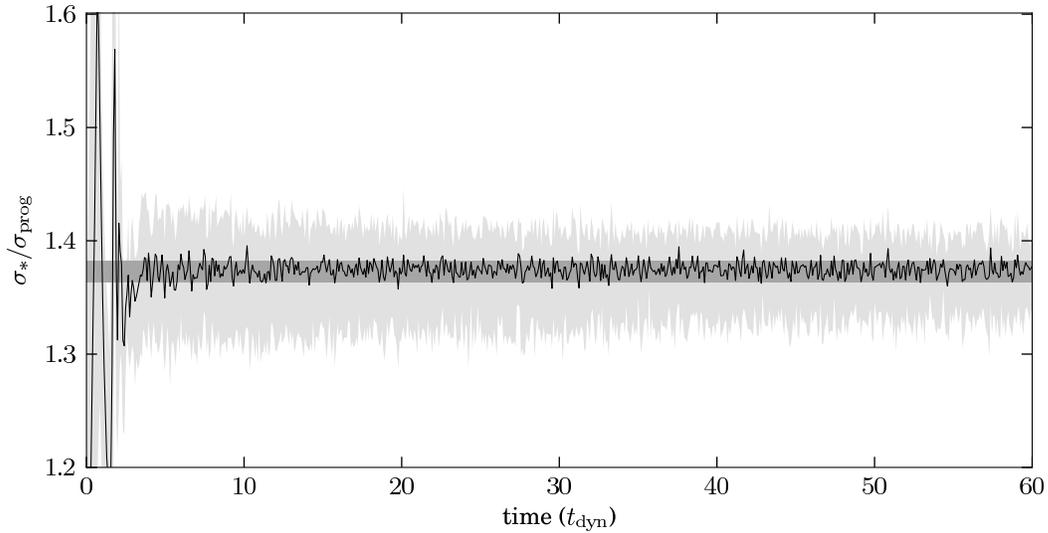}
     \caption{\label{intermediate-osc-long}\small The evolution of \sig\ with time during the ``Intermediate, long '' merger of Table \ref{params}. The black line, shaded region, and gray horizontal stripe are defined the same as in Figure \ref{headon-osc}. The position and thickness of the gray stripe are $(\mbox{m}\sigma_{*,{\rm final}}\pm\sigma_{\rm d,noise})/\sigma_{\rm prog} = 1.3730\pm0.0096$. The phase mixing stage ends at $t\approx11$ \tdyn.}
     \end{center}
\end{figure*}

In our Head-on and Intermediate merger simulations we observed many statistically significant variations in \sig\ during the phase mixing stage of evolution. These fluctuations can be seen in Figures \ref{headon-osc}, \ref{intermediate-osc}, \ref{headon-osc-long} and \ref{intermediate-osc-long}. We define the end of the phase mixing stage as the time after which the directional mean of \sig\ can be found within the 1$\sigma$ particle noise of the equilibrium value, $\sigma_{*,{\rm final}}\pm\sigma_{\rm d, noise}$, with 75\% confidence (note that the \textit{process} of phase mixing continues to occur ad infinitum---only the phase mixing \textit{stage} has an end). This is admittedly a somewhat arbitrary criterion. For the Head-on collision shown in Figures \ref{headon-osc-long}, the phase mixing stage ended at $\approx 17\, t_{\rm dyn}$. In the Intermediate simulation, shown in Figures \ref{intermediate-osc-long}, the end of the stage occurred at $\approx 11\, t_{\rm dyn}$. The Orbital decay simulation, shown in Figure \ref{decay-osc} did not have a clearly distinct phase mixing stage. In Figures \ref{headon-osc-long} and \ref{intermediate-osc-long} a small number of 2$\sigma$ fluctuations in \sig\ can be seen \textit{after} the stated end of the phase mixing stage. Relatively large fluctuations, such as these, occurred very infrequently after the end of the stage compared with their frequency before the end of the stage.

During the phase mixing stage, fluctuations in the mean value of \sig\ were accompanied by corresponding fluctuations in the minimum and maximum values of \sig. This finding is consistent with the work of \citet{merrall03}, which found that the virial ratios, $2T/W$ oscillated with time as the systems progressed toward dynamical equilibrium ($T$ and $W$ are the kinetic and potential energy, respectively). Similar oscillations were also discussed by \citet{funato1992a,funato1992b}. These oscillations appear to be due to the presence of small subsets of particles in the merging system with very similar orbital periods. The rate of phase mixing depends monotonically on the difference between periods, thus particles with very similar periods mix slowly, which explains why these smaller oscillations survive longer than the bulk oscillations in the initial oscillatory stage \citep{funato1992b}. \citet{merrall03} found that such oscillations diminished in the continuum limit, as $N\rightarrow\infty$, but they noted that the fluctuations are indeed statistically significant and represent a physical phenomenon that occurs in systems consisting of a \textit{finite} number of particles. 

Note that the only differences between our ``short'' and ``long'' simulations (from Table \ref{params}) were the number of particles, the gravitational softening length, and the \sig\ measurement interval, $\tau$. The initial orbital parameters, total mass, size, and shape of the density profile were identical. This enabled us to determine whether the fluctuations during the phase mixing stage differed for systems having the same macrostate, but different microstates. We compared the evolution of \sig\ during the interval shared by both the ``long'' and ``short'' simulations (i.e., the first 8 \tdyn\ of each simulation) and found that the variations in \sig\ were in perfect agreement within the noise limits and sampling frequency. Notably, every fluctuation in the ``long'' simulation was accompanied by a corresponding fluctuation in the ``short'' simulation.  Of course not every fluctuation present in the ``short'' simulations could be detected in the ``long'' simulations because the sampling frequency in the ``long'' simulations was a factor of 10 lower and the noise threshold was slightly higher. Furthermore, the pseudorandom number generator function that was used in the galaxy construction algorithm was initialized with a different random seed each time a new spheroid was constructed. This caused the particles to be arranged differently each time a new system was constructed. Therefore, the similarity between the ``short'' and ``long'' simulations implies that the statistically significant fluctuations observed during the first 8 \tdyn\ of the ``long'' simulations were evidently not artifacts of the detailed microstates of the particle systems. The fluctuations depended on the macrostate---the density profile, velocity distribution, and initial orbital parameters.

As noted previously, the phase mixing stage in the Head-on collision lasted longer than the phase mixing stage in the Intermediate collision. The Orbital decay merger lacked a distinct phase mixing stage.  It appears that the duration of the phase mixing stage is inversely related to the duration of the oscillation stage. The magnitude of the time derivative of the mean gravitational potential in the early stages of the merger may be responsible for this relationship, however much more analysis would be required to verify this hypothesis. The efficiency of the mixing process may also play a role in determining the duration of the phase mixing stage. It is clear that the merger with the largest angular momentum (Orbital decay) exhibited the most short-lived phase-mixing stage, while the merger with zero angular momentum (Head-on) exhibited the longest-lived phase mixing stage. The (differential) rotation due to the presence of angular momentum, combined with dynamical friction may have accelerated the mixing process.

\subsubsection*{Dynamical Equilibrium}

Once the phase mixing stage was complete, we observed no significant evolution in the mean value of \sig\ in the remnant systems during the course of the $\approx 290$ \tdyn\ that the simulations were allowed to evolve. We determined the relaxed values of the mass-weighted velocity dispersion ($\mbox{m}\sigma_{*,{\rm final}}$) by  computing the time average of \sig\ during the final $\approx 200$ \tdyn\ of the ``long'' simulations. The mean value remained approximately constant---only rarely exceeding the 1$\sigma$ noise limit.  For the Head-on merger,  $\mbox{m}\sigma_{*,{\rm final}}=1.54\pm0.01\,\,r_{\rm h}\,t_{\rm dyn}^{-1}$. The Intermediate and Orbital decay mergers had identical velocity dispersions within the measurement uncertainty, with $\mbox{m}\sigma_{*,{\rm final}}=1.469\pm0.010\,\,r_{\rm h}\,t_{\rm dyn}^{-1}$ and $\mbox{m}\sigma_{*,{\rm final}}=1.456\pm0.010\,\,r_{\rm h}\,t_{\rm dyn}^{-1}$ respectively. As expected from the initial conditions, the Head-on merger remnant exhibited no net rotation; it was entirely supported by pressure. The Intermediate and Orbital decay remnants were supported by both rotation and pressure.

We observed the same general remnant properties described by \citet{gonzalez2005} and \citet{villumsen1982}. In particular, for the Head-on collision remnant, the velocity dispersion was largest along the collision axis and smallest perpendicular to the axis.  For the rotating remnants, the mean velocity dispersion in the orbital plane of the collision was larger than the velocity dispersion perpendicular to the orbital plane.

\begin{figure*}
 \begin{center}
    \includegraphics[width=5.9in]{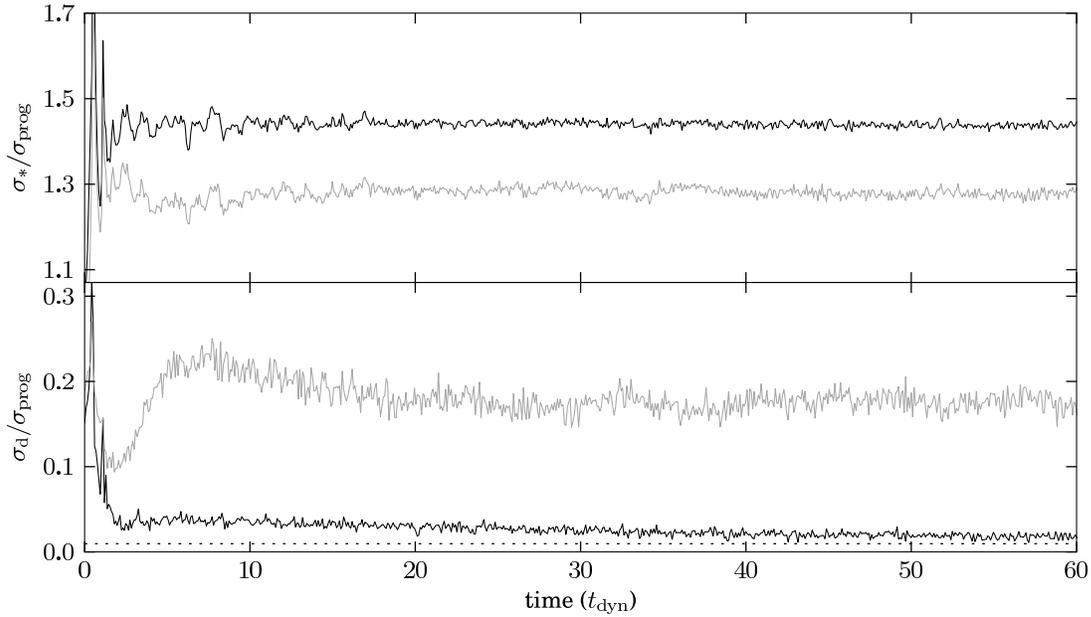}
    \caption{\label{headon-fluxstat}\small The evolution of \sig\ with time during the ``Head-on, long '' merger of Table \ref{params}.
 \textbf{Upper panel:}  The black and gray lines show the evolution of m\sig\ and f\sig\ with time, respectively. The flux-weighted quantity is systematically lower than its mass-weighted counterpart.
\textbf{Lower panel:} The black and gray lines show the dispersion in the mass-weighted and flux-weighted values of \sig\ for a set of $10^3$ random viewing directions (\sigd). This is effectively a measure of the anisotropy of the system. The lower dotted line shows the particle noise threshold which represents a lower limit on \sigd; a perfectly isotropic system composed of $7\times10^4$ particles would have a value of $\sigma_{\rm d}/\sigma_{\rm prog} = 0.0096$, which is the position of this line. From this plot, we can see that flux-based measurements would indicate that the system is much less isotropic than it actually is. This means that flux-based measurements of \sig\ from different viewing directions could differ significantly. The mass-weighted measurement shows that the system continued evolving toward a more isotropic state well after the end of the phase mixing stage.} 
 \end{center}
\end{figure*}

\begin{figure*}
 \begin{center}
    \includegraphics[width=5.9in]{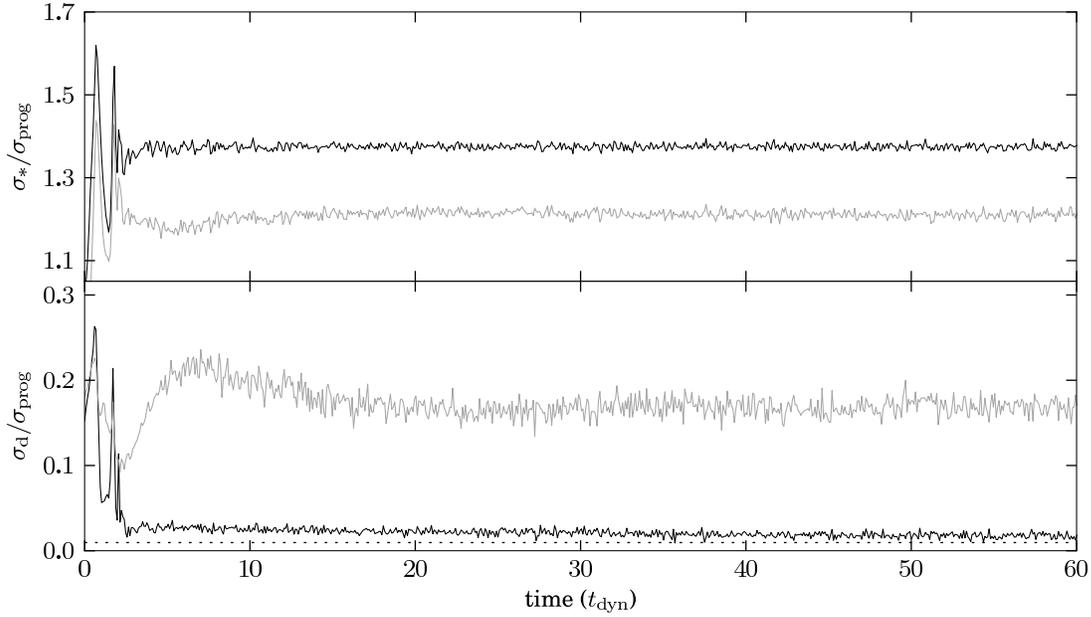}
    \caption{\label{intermediate-fluxstat}\small The evolution of \sig\ with time during the ``Intermediate, long '' merger of Table \ref{params}. See the caption of Figure \ref{headon-fluxstat} for a description of the plotted quantities. The effects of flux-weighting are the same as in the Head-on merger.} 
 \end{center}
\end{figure*}

\begin{figure*}
 \begin{center}
    \includegraphics[width=5.9in]{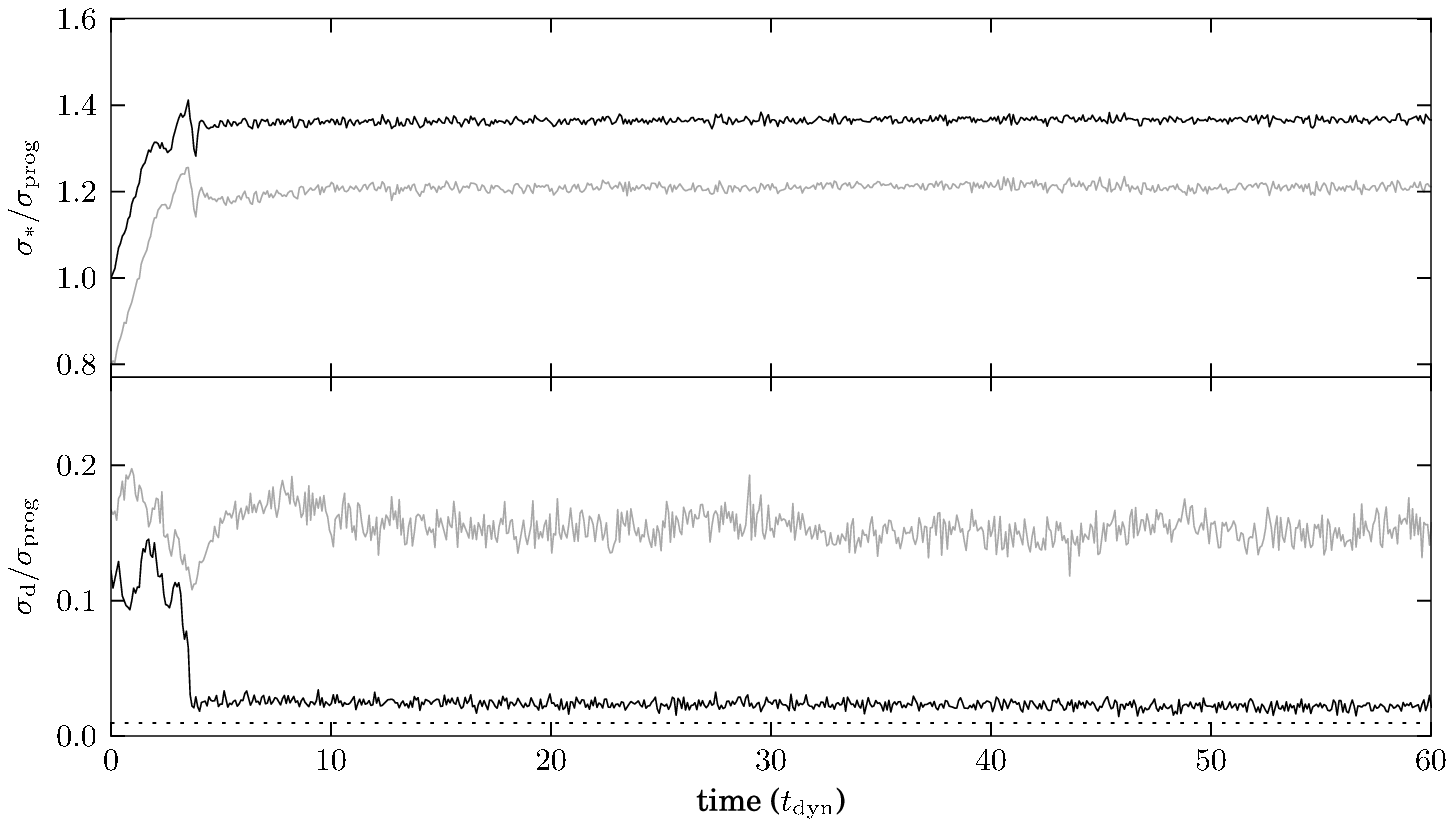}
    \caption{\label{decay-fluxstat}\small The evolution of \sig\ with time during the ``Orbital decay, long '' merger of Table \ref{params}. See the caption of Figure \ref{headon-fluxstat} for a description of the plotted quantities. Note that the remnant system shown here is less isotropic than the Head-on and Intermediate merger remnants. This is because the system is somewhat flattened due to rotation.} 
 \end{center}
\end{figure*}

Figures \ref{headon-fluxstat}--\ref{decay-fluxstat} show the evolution of m\sig\ and  f\sig\ along with their corresponding directional standard deviations m\sigd\ and f\sigd\ for our three merger simulations. While there was no evolution in the \textit{mean} of \sig, the directional distribution of \sig\ and the maximum value of \sig\ over the set of directions did evolve somewhat during the dynamical equilibrium stage.  This evolution is particularly evident in Figures \ref{headon-osc-long}, \ref{headon-fluxstat}, and \ref{intermediate-fluxstat}. The anisotropy (\sigd) and the maximum value of \sig\ decreased and eventually reached stable values a few tens of dynamical timescales after the end of the phase mixing stage. 

\subsection{Additional Statistics} 

Suppose a non-quiescent galaxy is observed and a careful measurement of \sig\ is made.  With what probability will the measured \sig\ fall within a specified range $\pm\Delta\sigma_*$ of the system's eventual equilibrium value of \sig, (i.e., the value of \sig\ nominally used in the \msig\ and FP relation studies)? In other words, what degree of scatter is expected when measuring \sig\ in a non-quiescent system and is there an offset in \sig\ during dynamically non-quiescent times? The results presented above provide the first pieces of information needed to answer these questions. For instance, in Figure \ref{headon-osc}, we see that, even in the most violent of the 1:1 mergers that we studied, the maximum value of \sig\ that would ever be measured is less than a factor of two greater than the final equilibrium value. Neglecting the effect of flux-weighting, the minimum value of \sig\ measured in an apparently coalesced system was about 70\% of the equilibrium value. We expect that \sig\ varies most significantly in 1:1 mergers, therefore this work suggests that a measurement of \sig\ made in a non-quiescent galaxy would fall between 70\% and 200\% of the quiescent value. Furthermore, a random  measurement of \sig\ is much more likely to fall near the equilibrium value than near an extremum; in order to obtain a measurement of \sig\ near an extremum, the system would have to be observed from a fortuitous viewing angle during a fairly short epoch. 

The standard deviation of \sig\ for the set of viewing angles in each merger is shown in the lower panels of Figures \ref{headon-fluxstat}--\ref{decay-fluxstat}. In these plots, we see that---in all three simulations---m\sigd\ was largest during the oscillatory stage with an absolute maximum value of about 25\% of the equilibrium value of \sig. After the oscillatory stage was complete, m\sigd\ was typically less than 3\% of the equilibrium value of \sig. The flux-weighted quantity, f\sigd\ was larger than m\sigd\ due to the anisotropy introduced by the slab of attenuating material.

We have not presented plots of the skewness or kurtosis of the directional distribution of \sig\ as a function of time because the large amount of noise present in these quantities prevented us from detecting any trends with time. There was, however, a clear difference between the mass-weighted and flux-weighted measurements.  These differences are summarized below. 
 
The skewness of the directional distribution of m\sig\ in the isolated progenitor system was $0.01\pm0.23$, which is consistent with zero, as expected from the manifest symmetry of the progenitor. Its flux-weighted counterpart was $-2.8\pm0.2$. The skewness of m\sig\ was $0.49\pm0.23$ for each of the three simulations during the oscillatory and early phase mixing stages of evolution, but the final quiescent values differed somewhat. The quiescent values of skewness for the Head-on, Intermediate, and Orbital decay merger remnants were respectively $0.07\pm0.22$, $-0.37\pm0.25$, and $-0.53\pm0.23 $. In each simulation, the skewness of the flux-weighted distribution was $-1.0\pm0.2$ during the oscillatory and early phase mixing stages and $-3.1\pm0.1$ for each of the remnant systems---consistent with the skewness of f\sig\ in the progenitors. Thus, the skewness of the flux-weighted velocity dispersion depended more strongly upon the dust geometry than the actual stellar dynamics.

Merger evolution had no measurable effect on the kurtosis of the directional distribution of m\sig. The kurtosis of m\sig\ was $2.5\pm 0.4$ for the progenitors and remained fixed at this value, within the limits of noise, for the duration of all three merger simulations. The kurtosis of f\sig\ in the progenitors, as well in each merger simulation, was consistently $11.6\pm1.9$. This elevated kurtosis indicates that much of the variability in the directional distribution of f\sig\ was due to extreme outliers, which is a direct result of the geometry of the attenuating slab.

\section{DISCUSSION AND CONCLUSIONS}

By analyzing the evolution of \sig\ during three collisionless, dissipationless, 1:1 mergers of spherically symmetric systems, we identified three primary stages of evolution. During the most dramatic, early stage of the merger, \sig\ undergoes large damped oscillations of increasing frequency. Following the oscillation stage, the value of \sig\ can fluctuate significantly in an apparently chaotic way for more than 10 dynamical timescales as the system becomes more mixed. We called this the phase mixing stage. The phase mixing stage is followed by the dynamical equilibrium stage during which the value of \sig\ remains essentially fixed, while the system evolves toward a final equilibrium state. Using the statistics computed during the collisions, we identified the extreme limits of \sig\ during such galaxy mergers and provided estimates of the scatter inherent in making measurements of \sig\ at random times from random viewing angles during a merger. The work of \citet{cox2006} hints that the evolution of \sig\ may be more complicated in mergers that contain a dissipative component (i.e., gas). If a similar analysis were performed on a large variety of more realistic dissipative galaxy mergers with varying mass ratios, gas fractions, Hubble types, and initial orbital parameters, it would be possible to predict the scatter inherent in making measurements of \sig\ in non-quiescent systems in general. 

By measuring \sig\ using a slit-based measurement method coupled with a toy model of dust attenuation, we found that the presence of dust in a galaxy can systematically decrease the flux-weighted value of \sig\ relative to the mass-weighted measurement. Furthermore, the distribution of dust can increase the apparent anisotropy of a system.  This increases the observed scatter in observational determinations of \sig. In order to understand how dust influences measurements of \sig\ in real galaxies, a more realistic model for dust attenuation is needed. We have begun a follow-up project that will use the radiative transfer code \textsc{Sunrise}, \citep{jonsson2006,jonsson2010} to create Doppler-broadened spectra of GADGET-2 \citep{springel2005} merger simulations at fixed time intervals. The synthetic spectra will be analyzed to determine the flux-weighted value of \sig. This will allow us to compute \sig\ in a way that is fully consistent with the method used by observers; not only will the \sig\ measurement be flux-based, but \sig\ will be obtained by fitting spectral line profiles as opposed to using direct particle data as in the present work. In addition to providing us with a better understanding of how \textit{dust} effects measurements of \sig, the follow-up project should also allow us investigate how \textit{star formation} influences the measurement of \sig. Rothberg \&\ Fischer recently reported a systematic discrepancy between values of \sig\ measured at different wavelengths, with lower \sig\ measured for longer wavelengths \citep{rothberg2010}. Their proposed explanation for this ``\sig\ discrepancy'' is as follows: stars for which \sig\ was measured using near-IR CO lines are young stars located in a dusty rotating gaseous disk whereas stars measured using the shorter wavelength Ca \textsc{iii} triplet were older. The young stars, having recently formed from collisional, dissipational, molecular gas, have lower velocity dispersion than the surrounding population of old stars because the clouds from which they formed had lower velocity dispersion than the older stellar population. These stars have not yet had time to mix with the older population and adopt the higher \sig. This explanation seems consistent with the observations of \citet{genzel2001} which found that the gas dynamics and stellar dynamics become decoupled during mergers. Our follow-up work should aid in understanding this ``sigma discrepancy''.

The merger simulations described in this work were intentionally kept simple in order to allow us to identify the most fundamental, purely dynamical aspects of the evolution of \sig\ during a merger. The systems were spherical, isotropic, non-rotating, and contained no gas nor dark matter. In spite of the simplicity, we observed nontrivial aspects of the evolution. In order to identify how each additional bit of complexity effects the evolution of \sig, we will compare these fundamental aspects of the evolution with the results of the more realistic simulations in our follow-up project. Even without performing further simulations, we can see that the time-scales and the set of possible initial orbital parameters will increase when we add a dark matter component. This is because systems of stars embedded in large dark matter halos are able to eventually merge as long as their parent halos interact sufficiently; the stellar systems themselves do not need to become superimposed during the earliest stages of the merger process because the earliest interactions primarily take place in the outer regions of the dark matter halos. The systems we modeled needed to \textit{start out} on a collision course in order to merge. Therefore, our simulations only represented the later stages of the overall merger process---beginning with the stage at which a collision between the stellar components was imminent.

\acknowledgements

We thank Gillian Wilson for her invaluable assistance. We are also grateful for the helpful feedback provided by Desika Narayanan, Mike Boylan-Kolchin, and the anonymous referee. Financial support for this work was provided by NASA through a grant from the Space Telescope Science Institute (Program AR-12626), which is operated by the Association of Universities for Research in Astronomy, Incorporated, under NASA contract NAS5-26555.  Additional support was provided by the National Science Foundation, under grant number AST 0507450. 

\appendix

\section{Analytic Results}\label{appendix}

The mass-weighted mean value of the radial velocity dispersion in a region of volume $V$ is given by 
\begin{equation}
 \mbox{m}\sigma_{*} = \sqrt{\mbox{m}\sigma_{\rm r}^2}
\end{equation}
with
\begin{equation}
 \mbox{m}\sigma_{\rm r}^2 = \frac{1}{M_V}\int_V \langle v_{\rm r}^2\rangle \rho \,dV
\end{equation}
\noindent
where $M_V$ is the mass within the volume $V$, $\rho$ is the mass density, and $\langle v_{\rm r}^2\rangle=\langle(\hat{\mathbf{r}}\cdot\mathbf{v_{\rm cm}})^2\rangle$ is the mean squared radial component of the velocity with respect to the center of mass of the system (i.e., the velocity in the zero momentum frame). For a spherically symmetric system, this can be written in terms of $r$ as
\begin{equation}
 \mbox{m}\sigma_{\rm r}^2(r) = \frac{4\pi}{M(r)}\int_0^r \langle v_{\rm r}^2\rangle \rho(r')r'^2 \,dr'\label{mmeanv}
\end{equation}
\noindent
where $M(r)$ is the mass enclosed within radius $r$. A system having an isotropic velocity dispersion tensor throughout will, by definition, have the same velocity dispersion in all directions at any arbitrary position $\mathbf{r}$ within the system. Thus, for an \textit{isotropic} system, we could measure the velocity dispersion along \textit{any} direction, compute the mass-weighted mean, and arrive at the same result for m\sig. In particular, if we choose the line-of-sight direction, $\hat{\mathbf{n}}$, then $\langle v_{\rm n}^2\rangle=\langle(\hat{\mathbf{n}}\cdot\mathbf{v_{\rm cm}})^2\rangle=\langle(\hat{\mathbf{r}}\cdot\mathbf{v_{\rm cm}})^2\rangle=\langle v_{\rm r}^2\rangle$. An expression for $\langle v_{\rm r}^2\rangle$ in an isotropic, non-rotating Hernquist profile is given in \citet{hernquist1990} and reproduced below:
\begin{equation}
 \langle v_{\rm r}^2\rangle  =  \frac{GM}{12D}\left\{\frac{12r(r+D)^3}{D^4}\ln\left(\frac{r+D}{r} \right) - \frac{r}{r+D}\left[25 + 52 \frac{r}{D} + 42\left( \frac{r}{D} \right)^2 + 12\left( \frac{r}{D} \right)^3 \right]\right\}\label{vr2units}
\end{equation}
\noindent
Adopting a system of units in which $G=D=M=1$, (\ref{vr2units}) becomes 
\begin{equation}
 \langle v_{\rm r}^2\rangle  =  r(r+1)^3\left[\ln\left(1+r^{-1}\right) - \frac{25 + 52 r + 42r^2 + 12r^3}{12(r+1)^4}\right]\label{vr2}
\end{equation}
\noindent
In these units, $ 1 \,r_{\rm h} = 1+\sqrt{2}$ and 
\begin{equation}
 1\, \frac{r_{\rm h}}{t_{\rm dyn}} = \sqrt{\frac{2}{\pi^2(1+\sqrt{2})}} \label{v_unit}
\end{equation}
\noindent
Upon substituting (\ref{vr2}) into (\ref{mmeanv}) with $r=r_{\rm h}$, evaluating the integral, computing the square root, and using (\ref{v_unit}) to express the result in terms of \rhalf\ and \tdyn, we arrive at the result, 
\begin{equation}
 \mbox{m}\sigma_* \approx 1.0035 \frac{r_{\rm h}}{t_{\rm dyn}}
\end{equation}
\noindent
Substituting $r=r_{\rm h}/2$ yields
\begin{equation}
 \mbox{m}\sigma_* \approx 1.0693 \frac{r_{\rm h}}{t_{\rm dyn}}
\end{equation}

\end{document}